\documentclass[aps,showpacs,superscriptaddress,amsfonts,onecolumn,prl]{revtex4-2}

\usepackage{graphicx}
\usepackage{array}
\usepackage{hyperref}
\usepackage{amsmath,amssymb}
\usepackage{float}
\usepackage{amsfonts}
\usepackage{ar}
\usepackage{xcolor}
\usepackage{physics}
\usepackage{soul}
\usepackage{siunitx}

\newcommand{\uni}{{\it uni1 }}
\newcommand{\wt}{{\it wt }}
\newcommand{\ptx}{{\it ptx1 }}

\begin{document}

\title{The role of hydrodynamics in the synchronisation of {\it Chlamydomonas} flagella}

\author{Luc Zorrilla}
\affiliation{Mediterranean Institute for Advanced Studies, IMEDEA, UIB-CSIC, Esporles, 07190, Spain}
\author{Antoine Allard}
\affiliation{Univ. Bordeaux, CNRS, LOMA, UMR 5798, F-33400 Talence, France}
\author{Krish Desai}
\email[]{Current affiliation: Department of Metabolism, Digestion and Reproduction, Imperial College London, UK.}
\affiliation{Department of Physics, University of Warwick, Gibbet Hill Road, Coventry CV4 7AL, UK}
\author{Marco Polin}
\email[]{email: mpolin@imedea.uib.csic}
\affiliation{Mediterranean Institute for Advanced Studies, IMEDEA, UIB-CSIC, Esporles, 07190, Spain}
\altaffiliation{Department of Physics, University of Warwick, Gibbet Hill Road, Coventry CV4 7AL, UK}

\date{\today}

\begin{abstract}
While hydrodynamic coupling has long been considered  essential for synchronisation of eukaryotic flagella, recent experiments  on the unicellular biflagellate model organism {\it Chlamydomonas} demonstrate that -at the single cell level- intracellular mechanical coupling is necessary for coordination. It is therefore unclear what role, if any, hydrodynamic forces actually play in the synchronisation of multiple flagella within individual cells, arguably the building block of large scale coordination. Here we address this question experimentally by transiently blocking  hydrodynamic coupling between the two flagella of single {\it Chlamydomonas}. 
Our results reveal that in wild type cells intracellularly-mediated forces are necessary and sufficient for flagellar synchronisation, with hydrodynamic coupling causing minimal changes in flagellar dynamics. However, fluid-mediated ciliary coupling is responsible for the extended periods of anti-phase synchronisation observed in a mutant with weaker intracellular coupling. 
At the single-cell level, therefore, flagellar coordination depends on a subtle balance between intracellular and extracellular forces.
\end{abstract}

\maketitle
Cilia and flagella, henceforth used interchangeably, are slender organelles ubiquitous in the eukaryotic world, where they support tasks ranging from mechanochemical sensing to regulate cell proliferation in renal ducts \cite{kathem13}, to pumping of cerebrospinal flow by ependymal cells \cite{mahuzier18} and establishment of planar cell polarity \cite{guirao10a,arata22}. 
Motile cilia combine thousands of molecular motors within a microtubule-based passive scaffold, the axoneme, rooted in an intracellular basal body \cite{klena22}. Their regular beating results from the motors' active stresses, the mechanical properties of the axoneme and the viscoelastic properties of the extracellular fluid \cite{leung21}. 

In groups of motile cilia, whether belonging to the same cell or to different ones, coordinated motion is a ubiquitous feature \cite{elgeti13}.
The metachronal waves observed at a large scale can increase transport efficiency in ciliated epithelia in a noise-dependent manner \cite{Ramirez-SanJuan2020} and improve swimming efficiency of ciliated microorganisms \cite{michelin10}. At the single-cell level, active modulation of flagellar synchrony can help microorganisms explore space \cite{polin09,poon25} and respond to environmental stimuli \cite{leptos23}. 
Fluid-mediated interactions between beating flagella have commonly been considered the cornerstone of flagellar synchronisation, usually coupled with either a modulation of the ciliary driving force \cite{meng21}, waveform compliance \cite{niedermayer08,brumley15} or a combination of the two \cite{maestro18}.  
Direct experiments with pairs of flagellated cells support this hypothesis: synchronisation develops spontaneously below a critical cell-to-cell separation as a result of hydrodynamic interactions \cite{brumley14}. Ciliary beating can also be entrained by an externally imposed flow, if sufficiently strong \cite{wei24}.

\begin{figure}[t]
\centering
\includegraphics[width=0.5\linewidth]{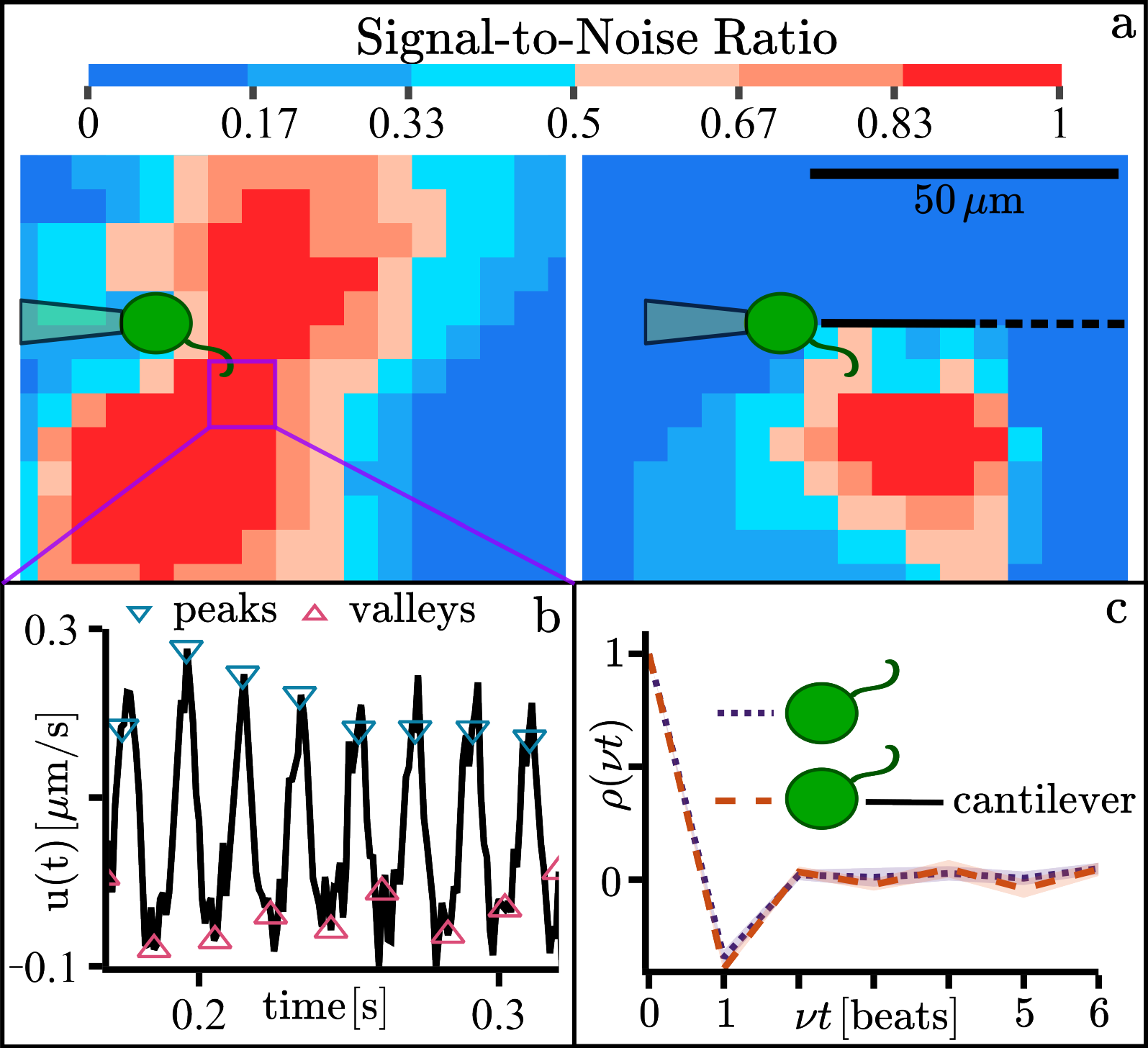}
\caption{\textbf{Flow field for uniflagellated mutants}. a) Signal to noise
ratio of the flagellar flow field without (left) and with the cantilever (right). 
b) Time-dependent flow field averaged across the highlighted area. Peaks and valleys
are used to define the Poincar\'{e} sections of the phase cycle. 
c) Autocorrelation of the normalised instantaneous periods between beats.}
\label{fig1}
\end{figure}

The focus on hydrodynamic coupling between flagella has been more recently upended by experiments on single cells. Studies on {\it Xenopus} multiciliated cells \cite{werner11} and {\it Paramecium} \cite{narematsu15} suggested that intracellular mechanical coupling between  flagellar basal bodies are critical for flagellar coordination. 
This was confirmed by experiments on the  {\it vfl3} mutant strain of the biflagellate unicellular green alga {\it Chlamydomonas reinhardtii} (CR), a major model system to study eukaryotic flagella \cite{quaranta15,wan16}. These mutants lack the intracellular fibres that, in the wild type, connect the two flagellar basal bodies \cite{wright83}. In their absence, flagella do not synchronise, showing that intracellular mechanical coupling is necessary for the coordination of CR flagella. 
The importance of direct basal body connections has also been shown in {\it Tetrahymena}, where cortical actin is required for correct basal body placement and ciliary coordination  \cite{soh22}.
The mechanism leading to flagellar coordination within a single cell has been explored theoretically through minimal models combining basal coupling and hydrodynamic interactions between flagella \cite{klindt17,liu18,man20,guo21}. These models provide useful insights on the different synchronisation regimes possible through the interplay of extracellular and intracellular coupling, but are all based on experiments that have probed only the effect of removing the latter.
Understanding the mechanism leading to flagellar coordination within a single cell requires us to be able to selectively remove hydrodynamic coupling between the flagella. So far this has proved challenging.

Here we use a tip-less cantilever to separate the flow fields generated by the two flagella of individual CR cells, allowing us to compare their behaviour with and without hydrodynamic coupling. While the standard in-phase (IP) synchrony is maintained in wild type ({\it wt}) regardless of the cantilever, the prolonged periods of anti-phase (AP) coordination displayed by the flagellar dominance mutant {\it pxt1} require hydrodynamic coupling. Our results provide novel critical insights into the role of hydrodynamics on the collective behaviour of eukaryotic flagella.

{\it Chlamydomonas reinhardtii} strains CC125 ({\it wt}, wild type), CC2894 ({\it ptx1}, flagellar dominance mutant) and CC1926 ({\it uni1}, uniflagellated) (\texttt{chlamycollection .org}) were grown in Tris-Acetate-Phosphate medium (TAP; \cite{gorman65}) within a diurnal growth chamber ($20^{\circ}\,$C; 14/10 day/night; $80\,\mu$mol/m$^2$s PAR). A small aliquot of exponentially-growing cells was then injected into a $4\,$mm-thick poly(dimethylsiloxane) observation chamber filled with a suspension of $1\,\mu$m carboxylate polystyrene tracer particles (Polysciences Europe, Germany) in TAP ($0.026\%$ solids). Individual cells (13 \uni; 8 {\it wt}; 14 {\it ptx1}) were captured on the tip of a glass micropipette hosted on a micromanipulator (PatchStar, Scientifica, UK) and realigned to beat their flagella along the focal plane of a $63\times$ water immersion objective (NA 1.0; Zeiss, Germany). The support section of a tip-less silicon nitride AFM cantilever (Quest R 20, NuNano, UK; $0.65\times30\times200\,\mu$m) was previously inserted into the top side of the chamber, allowing us to place the cantilever sideways in front of the cell (Fig.~\ref{fig1}a). Pre-treating the chamber with a $0.5\%$ w/v pluronic solution prevented bead adhesion to the cantilever.  The flagellar flow field was imaged in brightfield through a $610\,$nm long-pass filter and recorded at $500\,$fps ({\it wt},{\it ptx1}; $1000\,$fps {\it uni1})  (EoSens Cube6/mini1, Mikrotron, Germany). The local instantaneous flow field around each cell with/without the cantilever was estimated {\it via} Particle Image Velocimetry \cite{thielicke14}. 

\begin{figure}[t]
\centering
\includegraphics[width=0.4\linewidth]{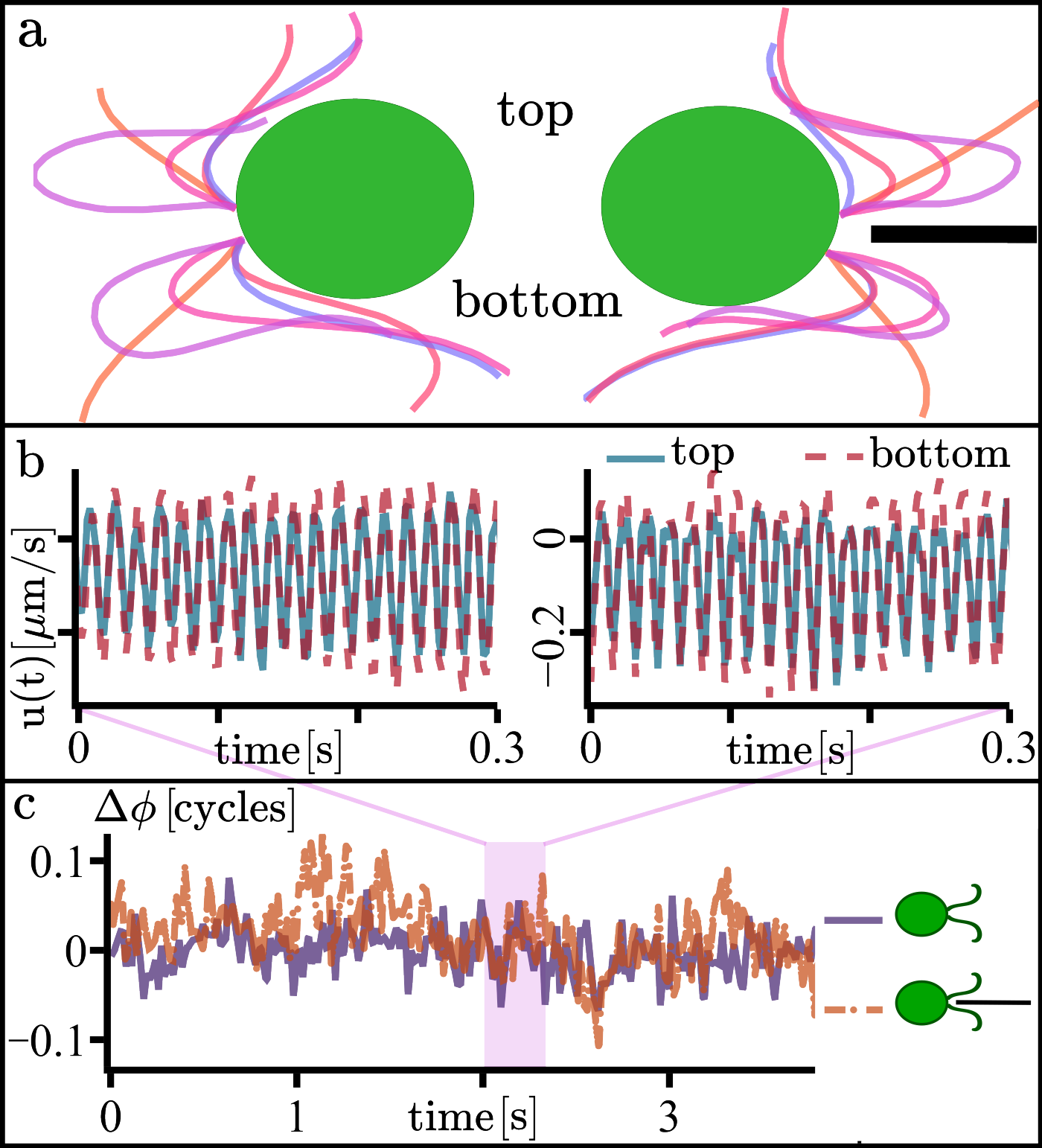}
\caption{\textbf{Flagellar synchronisation: bulk vs. cantilever ({\it wt}).} a) Stroboscopic traces of experimental flagellar waveforms without/with the cantilever. b) Longitudinal flagellar flow fields of top and bottom without (left) and with (right) the cantilever. c) Inter-flagellar phase difference without/with the cantilever.}
\label{fig2}
\end{figure}

\begin{figure}[t]
\centering
\includegraphics[width=0.5\linewidth]{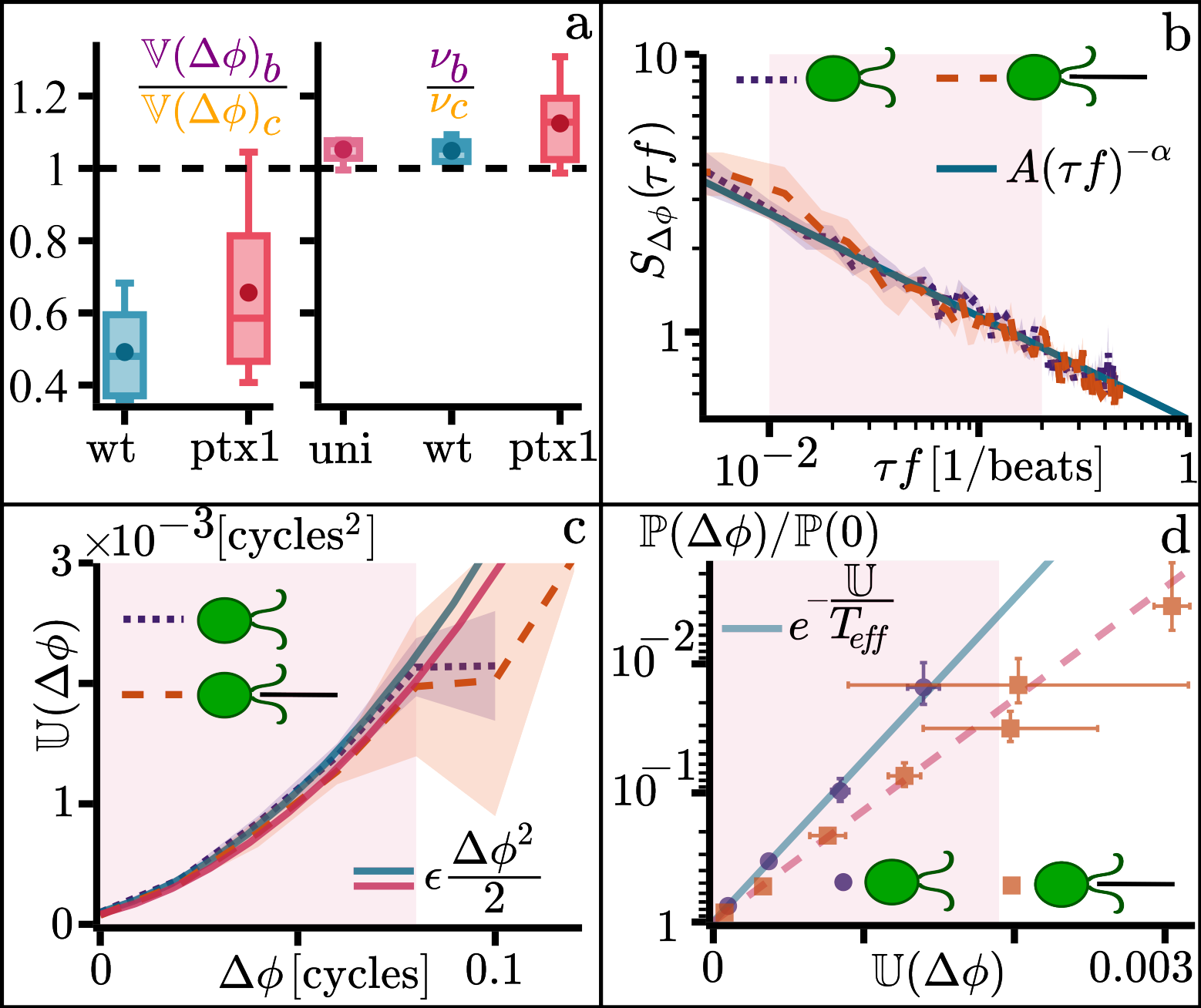}
\caption{\textbf{Inter-flagellar phase difference $\Delta\phi$: bulk vs. cantilever ({\it wt}).} a) Variance of $\Delta\phi$ ($\mathbb{V}(\Delta\phi)$) and mean beating frequency ($\nu$): ratio between bulk and cantilever values (subscripts $b$ and $c$ respectively).  b) Power spectrum $S_{\Delta\phi}$ of phase difference fluctuations during synchronous periods with (orange dashed) and without (purple dashed) the cantilever. Solid line: fit of the bulk case to the inverse power law $A/f^{\alpha}$. c) Effective potential $\mathbb{U}(\Delta\phi)$ with (orange dot-dashed) and without (purple dashed) the cantilever, and quadratic fits (pink and blue solid lines respectively).  d) Scatter plot of $\mathbb{P}(\Delta\phi)/\mathbb{P}(0)$ vs. $\mathbb{U}(\Delta\phi)$ and fit to $\exp(-\mathbb{U}/T_{\text{eff}})$ with (orange squares and dashed line) and without (purple disks and solid line) the cantilever. Shaded regions indicate the ranges of data used for the fits.}
\label{fig3}
\end{figure}

Figure~\ref{fig1}a characterises the local strength of the periodic flow from the single flagellum of {\it uni1}, as the normalised ratio between the excess spectral power around the beating frequency and the baseline power level from the noise \cite{supmat}. We see that an aptly placed cantilever confines flagellar flow to one side of the cell (configuration {\it ctl}; see also Movie~M1), confirming that this configuration can block inter-flagellar hydrodynamic coupling in biflagellate cells. The local oscillatory flow (Fig.~\ref{fig1}b) can also be used to estimate the beating phase $\phi(t)$ of each flagellum through Poincar\'e sectioning of the dynamics, as done previously \cite{leptos13,brumley14,pellicciotta20}. These phases can be used to compare, for each cell, characteristics of flagellar beating with and without the cantilever. As seen in Fig.~\ref{fig1}c the cantilever does not alter the autocorrelation of the instantaneous beating periods, which resembles that of a slightly underdamped oscillator as previously reported \cite{wan14}.
Having established that the cantilever prevents the flow generated by one flagellum from crossing over to the site of the other,  we now turn to \wt biflagellate cells and use the cantilever to prevent hydrodynamic coupling. Movies~M2,M3 show that, when placing the cantilever, flagellar synchronisation in \wt cells not only is preserved, but it is also recovered after a photoshock, induced here with a flash of blue light ($470\,$nm). Figure~\ref{fig2} shows that the presence of the cantilever does not significantly alter the waveform (Fig.~\ref{fig2}a), the longitudinal flow field close to the flagellum $\text{u}(t)$ (Fig.~\ref{fig2}b), or the mean around which the inter-flagellar phase difference $\Delta\phi(t)$ fluctuates (Fig.~\ref{fig2}c). This proves that flagellar synchronisation in \wt does not {\it per se} require hydrodynamic coupling: intracellular coupling is both necessary {\it and sufficient} to establish and maintain the synchronised motion that is usually observed. Yet, one would be naive to conclude that hydrodynamic forces have no measurable consequences in this system. 
Figure~\ref{fig3}a shows that all of the strains we tested display a minute increase in beating frequency ($\sim5\%$) without the cantilever (a configuration henceforth referred to as {\it bulk}), probably due to a slightly lower flagellar drag coefficient (for a table collating all numerical values see \cite{supmat}). A more pronounced difference, however, can be found in the fluctuations of the inter-flagellar phase difference $\Delta\phi$ during synchrony. Its variance $\mathbb{V}(\Delta\phi)$ increases by a factor of $2.10 \pm 0.23$ when the cantilever is placed between the flagella ($1.56 \pm 0.15$ for {\it ptx1}). 
The change is not due to a difference in the dynamics of inter-flagellar phase fluctuations. Figure~\ref{fig3}b shows that both with and without the cantilever, the power spectrum $S_{\Delta\phi}$ of $\Delta\phi(t)$ follows the same curve. This is compatible with an inverse power law $\sim 1/f^{\alpha}$ with $\alpha_{\text{bulk}}^{wt}=0.37\pm0.02$ and $\alpha_{\text{ctl}}^{wt}=0.37\pm0.03$. Scale-free power spectra indicate long-range fractal-like temporal correlations in the fluctuations, and are a common phenomenon in nature, where they have been proposed to signal systems working at criticality \cite{gisiger01}. In CC125, temporal fluctuations in the flagellar beating period, as opposed to $\Delta\phi$, have also been reported to display scale-free fluctuations \cite{wan14}. In line with past works \cite{niedermayer08,brumley15,Ramirez-SanJuan2020} it is useful to conceptualise fluctuations in $\Delta\phi$ as a  stochastic process confined within a one-dimensional potential $\mathbb{U}(\Delta\phi)$. This implies that the force $-\partial \mathbb{U}(\Delta\phi)$ that the system feels at $\Delta\phi$ can be estimated as the expectation value  $\text{E}[\Delta\phi(t+\delta t)-\Delta\phi(t)|\Delta\phi(t)=\Delta\phi]/\delta t$ and then integrated to give the effective potential. Figure~\ref{fig3}c shows that $\mathbb{U}(\Delta\phi)\simeq \epsilon \Delta\phi^2/2$ both with and without the cantilever. The effective coupling constants $\epsilon^{\it wt}_{\text{bulk}} = 0.434 \pm 0.04$ and  $\epsilon^{\it wt}_{\text{ctl}}=0.402\pm0.06$ are not significantly different, in line with the fact that the cell maintains its in-phase (IP) beating regardless of the configuration. There is, however, a sizeable difference in the magnitude of the fluctuations within these potentials. This can be estimated by calculating the probability distribution function $\mathbb{P}(\Delta\phi)$ of the excursions of the phase difference with respect to the plateau value during IP, and comparing $\mathbb{U}(\Delta\phi)$ to $-\log\mathbb{P}$. Figure~\ref{fig3}d) shows that the two align well, implying that -despite their memory- fluctuations in $\Delta\phi$ are ergodic with respect to a simple Boltzmann-like measure. The slope of the linear fit can then be interpreted as the inverse of an effective temperature $T_{\text{eff}}$, characterising the noise magnitude. We obtain $T_{\text{eff, bulk}}^{wt} = (3.46 \pm0.11)\times 10^{-4}$, $T_{\text{eff, ctl}}^{wt} = (5.56\pm0.29)\times 10^{-4}$. The change in variance in Fig.~\ref{fig3}a can then be understood as a change in the ratio $\epsilon/T_{\text{eff}}$, where the main contribution comes from an increase in the magnitude of intrinsic fluctuations ($(\epsilon^{\it wt}_{\text{ctl}}T_{\text{eff, bulk}}^{wt})/(\epsilon^{\it wt}_{\text{bulk}}T_{\text{eff, ctl}}^{wt})=0.64\pm0.3$). 

Flagella from \wt cells, termed {\it cis} and {\it trans} depending on the position with respect to the cell's eyespot, are not identical organelles. They differ in structure \cite{dutcher19}, intrinsic frequency ($\sim 30\%$ \cite{kamiya87}) and coupling \cite{wei24}, causing the {\it cis} flagellum to dominate the dynamics of the pair. A conceptually simpler system is provided by {\it ptx1}, a mutant strain which putatively possesses two identical flagella of {\it trans} type \cite{huang82}. We previously reported that, differently from {\it wt}, pipette-held \ptx cells can display extended periods of either in-phase (IP) or anti-phase (AP) beating \cite{leptos13}, the latter being the state favoured by hydrodynamic interactions \cite{brumley14}. 
\begin{figure}[t]
\centering
\includegraphics[width=0.5\linewidth]{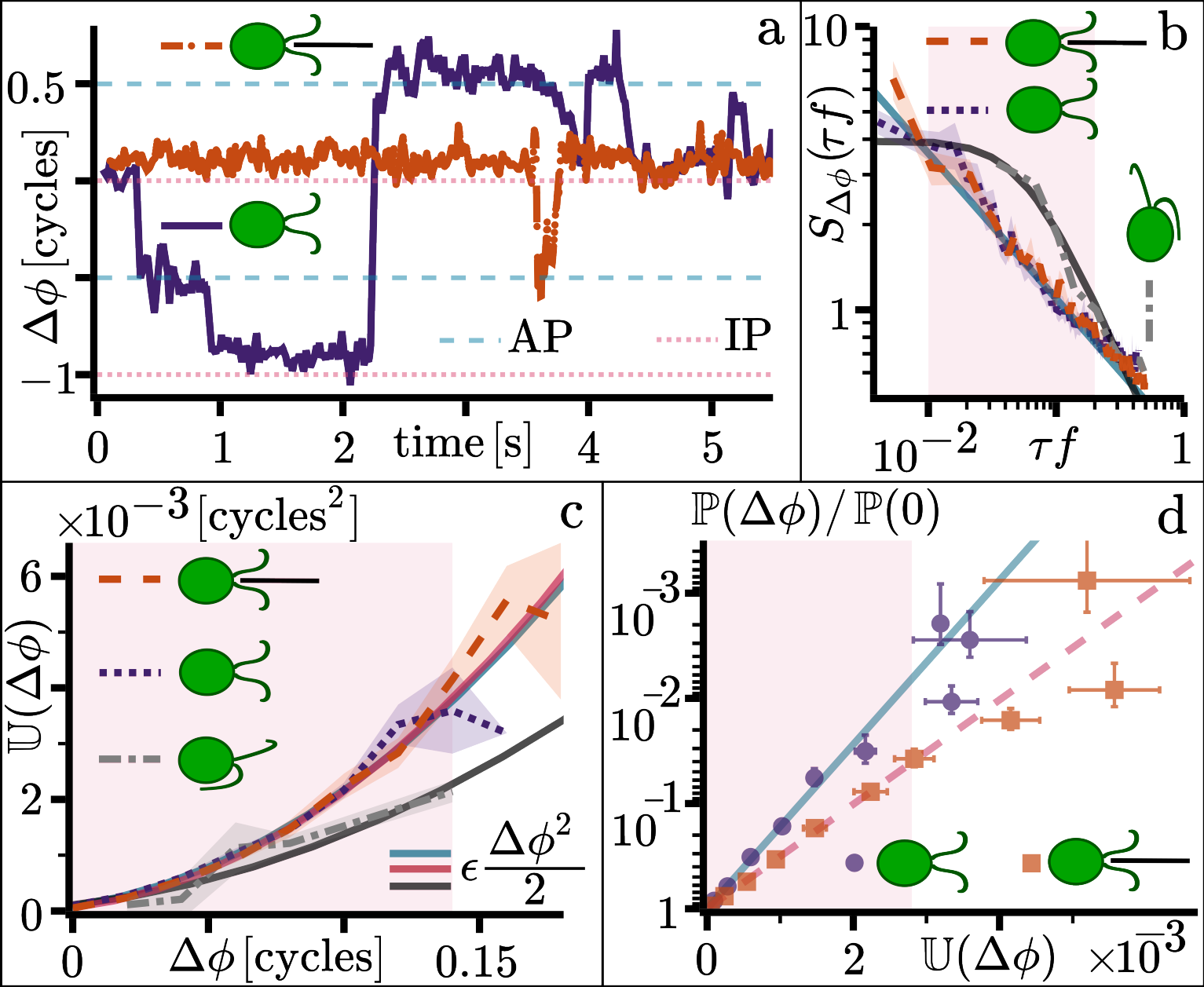}
\caption{\textbf{Inter-flagellar phase difference: bulk vs. cantilever ({\it ptx1}).}  a) Representative dynamics of $\Delta\phi$ without (purple solid line) and with (dot-dashed orange line) the cantilever. Integer/half integer values correspond to IP and AP synchronisation respectively. b) Power spectrum $S_{\Delta\phi}$ of phase difference fluctuations during synchronous periods with (IP: orange dashed) and without (IP: purple dotted; AP: pink dot-dashed) the cantilever. Solid lines: fit of bulk IP to an inverse power law $A/f^{\alpha}$ (blue) and of AP to a Lorentzian (pink). c) Effective potentials $\mathbb{U}(\Delta\phi)$: IP with (orange dashed) and without (purple dotted) the cantilever; AP (pink dot-dashed). Solid lines: corresponding quadratic fits (orange, blue and pink respectively).  d) Scatter plot of $\mathbb{P}(\Delta\phi)/\mathbb{P}(0)$ vs. $\mathbb{U}(\Delta\phi)$ and fit to $\exp(-\mathbb{U}/T_{\text{eff}})$ with (orange squares and dashed line) and without (purple disks and solid line) the cantilever. Shaded regions indicate the ranges of data used for the fits.}
\label{fig4}
\end{figure}
Figure~\ref{fig4} compares flagellar behaviour in \ptx with and without the cantilever. It shows that, while cells fluctuate between AP and IP periods in bulk, AP disappears when the cantilever is in place (fig.~\ref{fig4}a). Instead, IP synchrony is punctuated by sporadic short periods of large $\Delta\phi$ deviations accompanied by the beating frequency increase characteristic of AP ($\bar{\nu}=(77.0\pm 4)\,$Hz {\it vs} $\bar{\nu}_{AP}=(82.2\pm 5)\,$Hz and $\bar{\nu}_{IP}=(58.4\pm 2)\,$Hz in bulk) but never lasting more than $200\,$ms. AP synchronisation, therefore, requires inter-flagellar hydrodynamic coupling. Comparing the IP periods with and without the cantilever, we recover a similar behaviour as observed in {\it wt}. Cells in bulk have a slightly higher beating frequency ($12 \pm 3 \%$) and lower variance $\mathbb{V}(\Delta\phi)$ ($64 \pm 6 \%$) (fig.~\ref{fig3}a). Their $\Delta\phi$ fluctuations are characterised by compatible inverse-power-law-type power spectra $S_{\Delta\phi}(f)\sim 1/f^{\alpha}$ with $\alpha_{\text{bulk}}^{ptx1}=0.49\pm0.03$ and $\alpha_{\text{ctl}}^{ptx1}=0.48\pm0.02$. These are slightly larger than for {\it wt}, signalling a process with a somewhat shorter memory. 
As it is the case for {\it wt}, the deterministic part of the fluctuations dynamics of $\Delta\phi$ in \ptx is well captured by a quadratic effective potential, with coupling constants $\epsilon^{\it ptx1}_{\text{bulk}}=0.275\pm0.020$ and  $\epsilon^{\it ptx1}_{\text{ctl}}=0.299\pm0.024$ (fig.~\ref{fig4}c). These are not significantly different between them, but are only $\sim 69\%$ of the corresponding values in \wt, suggesting a weaker intracellular coupling between the two identical flagella of \ptx. Comparison between $\mathbb{U}(\Delta\phi)$ and $\mathbb{P}(\Delta\phi)$ shows that the fluctuations in $\Delta\phi$ are again compatible with a simple Boltzmann-like measure  (fig.~\ref{fig4}d).
This can be characterised by effective temperatures $T_{\text{eff, bulk}}^{ptx1} = (5.56\pm0.29)\times 10^{-4}$, $T_{\text{eff, ctl}}^{ptx1} = (8.75\pm0.16)\times 10^{-4}$, which are significantly larger ($\sim 167\%$) than for the \wt case. Once again, the change in variance in fig.~\ref{fig3}a can be understood as a change in the ratio $\epsilon/T_{\text{eff}}$, driven by a variation in $T_{\text{eff}}$  ($(\epsilon^{\it ptx1}_{\text{ctl}}T_{\text{eff, bulk}}^{ptx1})/(\epsilon^{\it ptx1}_{\text{bulk}}T_{\text{eff, ctl}}^{ptx1})=0.69\pm0.3$).

As mentioned earlier, the IP state in \ptx is accompanied -in bulk- by periods of AP synchronisation. Although the lack of AP states with the cantilever prevents a comparative analysis as done for IP, it is still instructive to analyse AP synchrony by itself. This reveals both quantitative and qualitative differences between AP and IP. Figure~\ref{fig4}c shows that the effective potential $\mathbb{U}(\Delta\phi)$ for AP is significantly weaker than for the IP of either \ptx or {\it wt}. The effective inter-flagellar coupling $\epsilon^{\it ptx1}_{\text{bulk,AP}}=0.140\pm0.034$  has a magnitude which is only $\sim 51\%$ of the corresponding IP values in {\it ptx1}. 
Note that the coupling constants for IP and AP states appear with opposite signs within the standard stochastic Adler-model used frequently in the past to rationalise the dynamics of $\Delta\phi$  \cite{liu18}.
Furthermore, $\Delta\phi$ fluctuations within this effective potential follow a power spectral density (fig.~\ref{fig4}b) which -surprisingly- is compatible with the Lorentzian decay predicted for a standard Ornstein-Uhlenbeck process, rather than the scale-free power law observed for IP in both \ptx and \wt (see \cite{supmat} for the corresponding autocorrelation functions). 

The dynamic of the inter-flagellar phase difference is a window into the mechanism responsible for synchronisation. Altogether, the differences in dynamics between IP ({\it wt}, {\it ptx1}) and AP ({\it ptx1}) suggest that these states stem from fundamentally different processes. 
On the one hand, the IP results presented here show unequivocally that this mode of flagellar synchronisation is independent of hydrodynamic coupling. 
Building upon an existing body of evidence \cite{werner11,narematsu15,quaranta15,wan16,soh22} we can therefore conclude that, in CR, IP stems exclusively from intracellular coupling, most likely realised by the fibrous connections between the flagellar basal bodies. 
The fluctuation dynamics observed in IP states ({\it wt}, {\it ptx1}) suggest a non-trivial stochastic process with a long memory. This might then be mediated by changes in the fibres' stiffness, possibly caused by local [Ca$^{2+}$] fluctuations \cite{wheeler08,collingridge13}. 
On the other hand, we see that AP requires hydrodynamic interactions, is associated with an effective coupling of reduced magnitude -and of opposite sign than IP- as well as a qualitatively different fluctuation dynamic which is conceptually simpler than in the IP case. We propose that these observations support the hypothesis of a fundamentally hydrodynamic origin of AP synchrony in {\it ptx1}, in line with reports on synchrony in pairs of nearby cells \cite{brumley15,wan16}. 
Previous theoretical investigations demonstrated the possibility to realise both AP and IP states with only hydrodynamic coupling \cite{man20}, or only through basal connections \cite{liu18,guo21}, or combining both \cite{klindt17,liu18}. In these models, areas of stable IP and AP in parameter space are separated by transition regions characterised by bi-stability and hysteresis, similarly to what is encountered in a subcritical pitchfork bifurcation \cite{strogatz15}.
Taken together with the observation of short large-magnitude lapses in IP synchrony when the cantilever is present, this suggests a picture where fluctuations of intracellular coupling in \ptx cause it to move outside the region of stable IP synchrony.
In bulk, hydrodynamics can then stabilise the system to AP. In order to revert to IP synchrony, then, the intracellular coupling will need to recover enough to overcome the hysteresis intrinsic within the bistable region, leading to prolonged periods of AP.
In turn, the dependence of AP on hydrodynamic coupling suggests that the presence of the cantilever should modify the stability profile in parameter space, drastically reducing both the AP region and the width of the AP-IP transition. This would then preclude the AP-like fluctuations observed in {\it ptx1} with the cantilever to extend into fully-fledged intervals of AP synchrony, as observed in experiments. At the same time, the existence of these fluctuations -lasting up to $\sim10$ beats- hints at the potential for a weak metastability induced solely by intracellular coupling in {\it ptx1}, as predicted by some models \cite{liu18,guo21}.
It would be interesting in the future to extend these studies by modulating the level of hydrodynamic interaction through partial insertion of the cantilever.
In the {\it wt} case, even without considering the effect of coupling asymmetries \cite{wei24}, the lack of AP synchronisation suggests that the intracellular coupling is large enough to guarantee that the system will fluctuate well within the stable IP phase. Indeed, this is supported by our estimate of the effective coupling strength, which is  $\sim 1.5\times$ the IP value of {\it ptx1}.
Besides its effects on the synchronisation state, the cantilever influences also the spread of phase difference fluctuations. Measured as an increase in $T_{\text{eff}}$, this suggests some level of flagellar sensing of the cantilever. 
{\it Chlamydomonas} flagella are decorated by mastigonemes, tiny transversal filaments $\sim 0.8\,\mu$m long \cite{wang23}. They do not play a direct role in motility \cite{amador20}, but are tethered to mechanosensitive proteins \cite{dai24}. We therefore surmise that the increase in $T_{\text{eff}}$ is the result of mastigoneme-mediated mechanosensitivity, an effect that would widen our current understanding of the link between mechanosensitivity and flagellar dynamics \cite{oshima23}. 
This will be probed further with targeted experiments using mutants lacking mastigonemes and/or mechanosensitive channels.

\begin{acknowledgements}
We thank Douglas Brumley for insightful discussions. The authors acknowledge support of the European Training Network PHYMOT (Horizon 2020 Marie Sk\l odowska-Curie grant agreement No 955910; L.Z. and M.P.); the Leverhulme Trust (grant RPG-2018-345; A.A. and M.P.); MICINN (grant PID2019-104232B-I00/AEI; M.P.). M.P. acknowledges the fact that IMEDEA is an accredited `Mar\'ia de Maeztu Excellence Unit' (grant CEX2021-001198, funded by MCIN/AEI/10.13039/ 501100011033). A.A. acknowledge support from the SMR Department and CNRS Tremplin@INP.
\end{acknowledgements}

\bibliography{Zorrilla25_flag_synch_bib}
\bibliographystyle{apsrev4-2}

\section{Supplementary plots}
\begin{figure}[h]
\centering
\includegraphics[width = 0.5\textwidth]{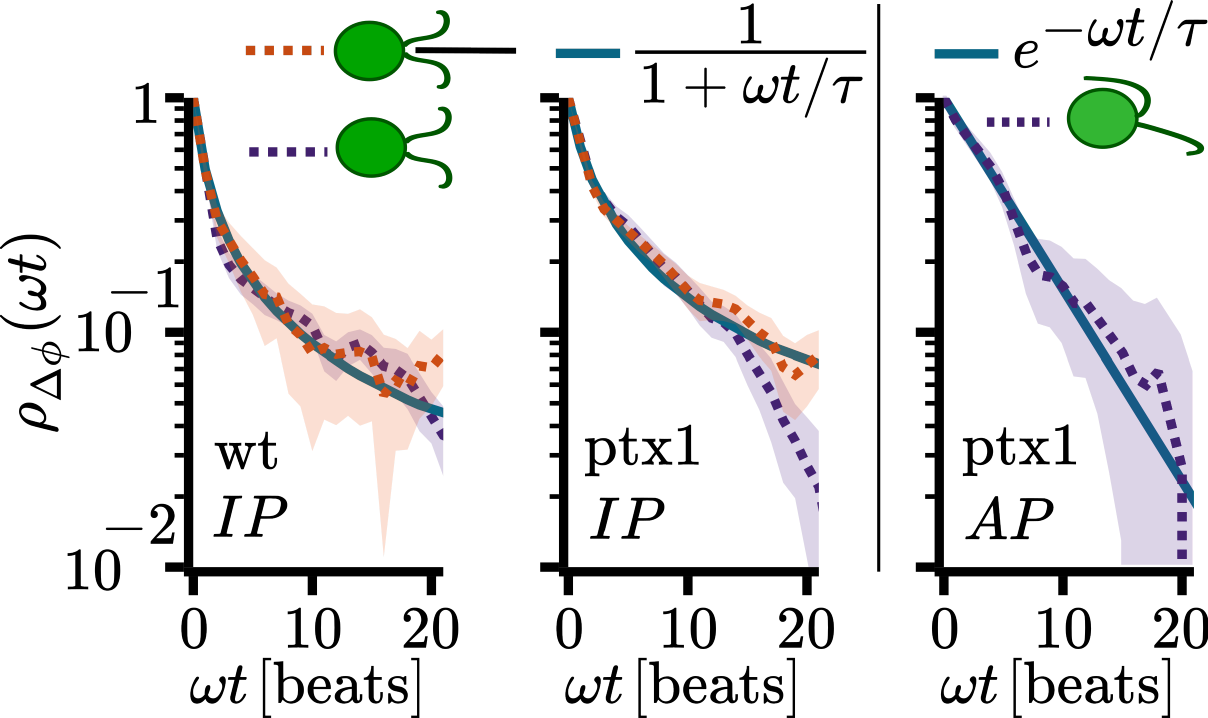}
\caption{Autocorrelation of the phase difference for different mutants and types of synchronization, with fits made on the whole visible range of data.  A homographic function is fitted for In-Phase synchrony, and only the fit to the bulk case is shown. An exponential function is fitted to Anti-Phase synchrony.}
\label{fig:rho_dphi_all}
\end{figure}
\begin{figure}[h]
\centering
\includegraphics[width = 0.3\textwidth]{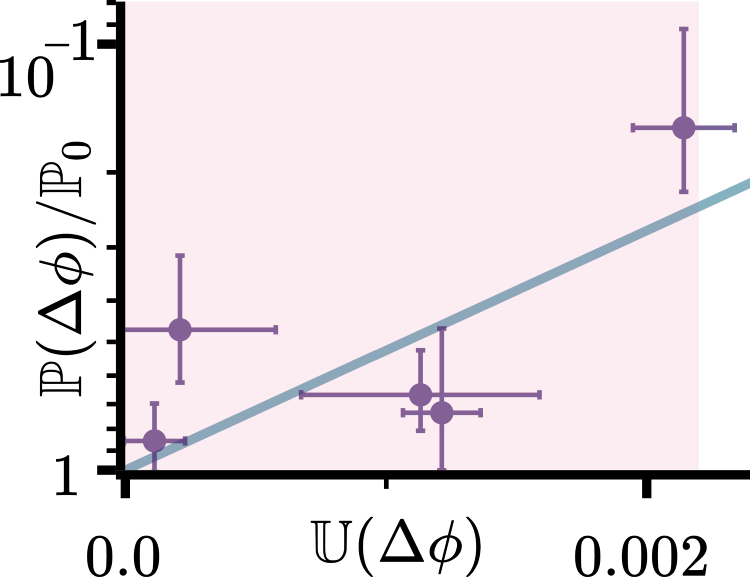}
\caption{Probability distribution function against potential, for \ptx cells in the AP synchronized regime, without cantilever. A fit to a Boltzmann distribution is made.}
\label{fig:boltzmann_plot_AP}
\end{figure}

\section{Supplementary Movies}
\begin{itemize}
\item Movie S1. Micropipette-held uniflagellated cell (CC1926) with/without the cantilever (same cell). Movie frame rate: 1000fps.
\item Movie S2. Micropipette-held wild type cell (CC125) with/without the cantilever (same cell). Movie frame rate: 500fps.
\item Movie S3. Photo-shock and recovery of a micropipette-held wild type cell (CC125) with the cantilever. The photo-shock was elicited with a sudden pulse from a $470\,$nm blue LED (M470L2, Thorlabs). Movie frame rate: 500fps.
\end{itemize}

\section{Evaluation of the strength of the periodic flow}
\begin{figure}[t]
  \centering
  \includegraphics[width = \textwidth]{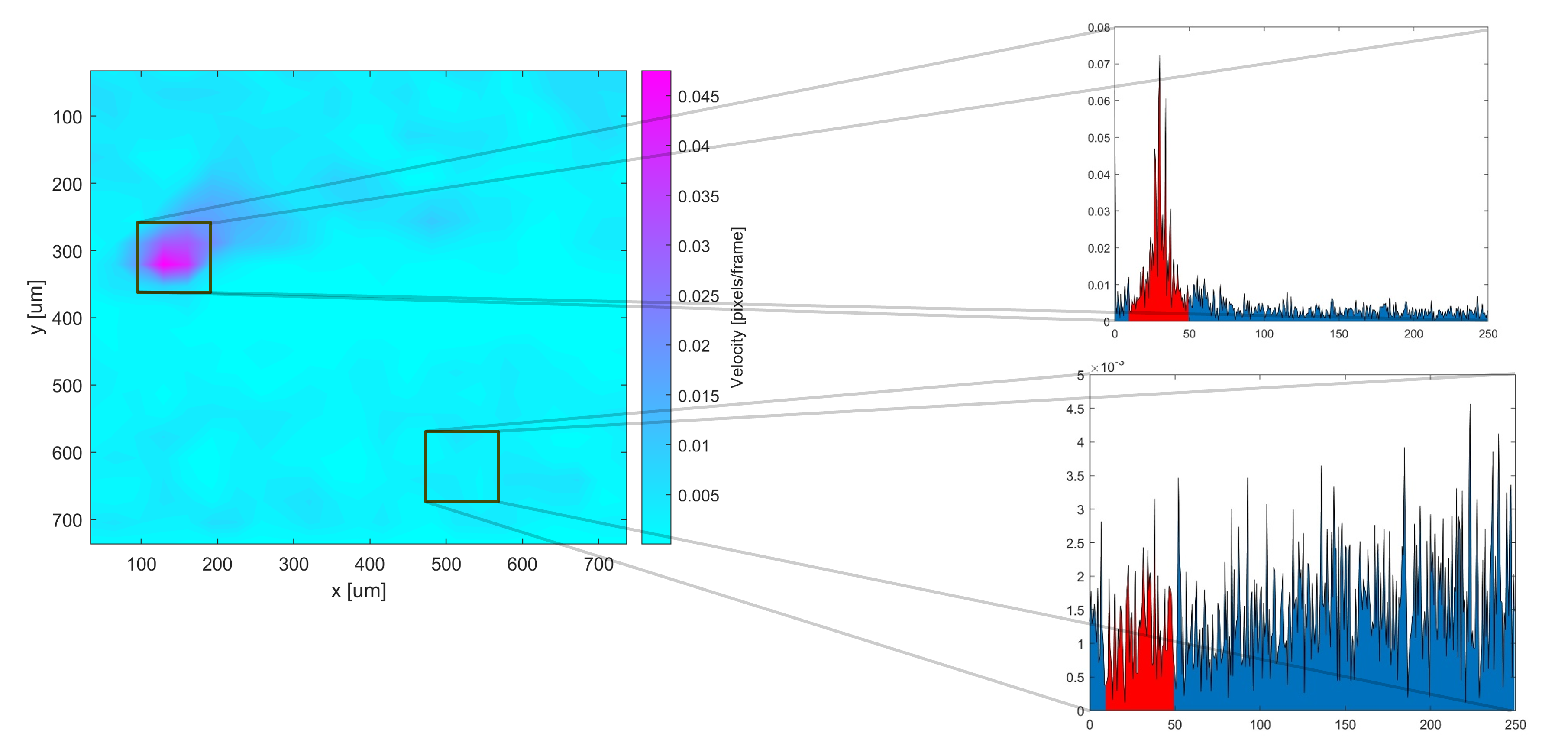}
  \caption{Constructing a Signal-to-Noise Ratio (SNR). (Left) An example time-averaged flow field retrieved with Particle-Image Velocimetry performed on a uniflagellated cell. (Right) Fourier transforms of the temporal flow at flagellum position (up) and far from the cell (down). The red color denotes the area under the curve used for SNR calculations.}
  \label{fig:uni_average_flow_field}
\end{figure}

We use PIV to estimate the instantaneous flow field $(u(\mathbf{x},t),v(\mathbf{x},t))$ on a $32\times32\,$pxl grid ($\sim 5.632\times 5.632\,\mu$m). At each point $\mathbf{x}$ along the grid we calculate the Fourier transform $\tilde{u}(\mathbf{x},\nu)$ of the $y$-component of the flow field $u(\mathbf{x},t)$. When present, the signature of flagellar beating is clearly visible as a peak in the power spectrum $|\tilde{u}(\mathbf{x},\nu)|^2$ at the beating frequency $\nu_0$ (see Fig.~\ref{fig:uni_average_flow_field}). We then calculate the signal $S(\mathbf{x})$ as the integral $$S_{u}(\mathbf{x})= \int_{\nu_0-\Delta\nu_0}^{\nu_0+\Delta\nu_0}|\tilde{u}(\mathbf{x},\nu)|^2d\nu,$$ where $\nu_0$ and $\Delta\nu_0$ are respectively the mean and the standard deviation of the instantaneous beating frequency (calculated from the time series of peak-to-peak intervals, see fig.1b of the main manuscript). We then perform a linear fit $P_{\text{lin}}(\mathbf{x},\nu)$ of the part of the power spectrum outside of the interval $[\nu_0-\Delta\nu_0,\nu_0+\Delta\nu_0]$ and use it to calculate $$S_{\text{lin}}(\mathbf{x})= \int_{\nu_0-\Delta \nu_0}^{\nu_0+\Delta\nu_0}P_{\text{lin}}(\mathbf{x},\nu)d\nu.$$
This estimates the power that would have been in $[\nu_0-\Delta\nu_0,\nu_0+\Delta\nu_0]$ in absence of a beating flagellum. The signal-to-noise ratio $$SNR(\mathbf{x}):=\frac{S_{u}(\mathbf{x})}{S_{\text{lin}}(\mathbf{x})}\frac{1}{\text{max}_{\mathbf{x}}(S_{u}(\mathbf{x})/S_{\text{lin}}(\mathbf{x}))}.$$
For sake of clarity, in fig.1 the signal $SNR(\mathbf{x})$ is binned in 6 bins for which the lowest corresponds to the noise level.

\section{Collated numerical results}
\begin{table}[t]
    \centering
    \begin{tabular}{|c|c|c|c|c|}
        \hline
        & CC1926 ({\it uni1}) & CC125 ({\it wt, IP}) & CC2894 ({\it ptx1}, IP) & CC2894 ({\it ptx1}, AP) \\ 
        \hline
        $\nu_b$ & $50.0 \pm 3.4$ & $52.8 \pm 2.9$ & $58.4 \pm 1.6$ & $82.2 \pm 4.9$ \\ \hline
        $\nu_c$ & $50.1 \pm 2.1$ & $50.4 \pm 2.9$ & $51.7 \pm 2.5$ & $77.0 \pm 4.0$ \\ \hline 
        $\left<\nu_b / \nu_c \right>$ & $1.054 \pm 0.018$ & $1.048 \pm 0.012$ & $1.124 \pm 0.033$ & $1.093 \pm 0.075$ \\ \hline        
        $\mathbb{V}(\Delta\phi)_b$ & N/A & $(0.66 \pm 0.10)\times 10^{-3}$ & $(1.86 \pm 0.22)\times 10^{-3}$ & $(95.4 \pm 37.4)\times 10^{-3}$ \\ \hline
        $\mathbb{V}(\Delta\phi)_c $ & N/A & $(1.55 \pm 0.36)\times 10^{-3}$ & $(2.91 \pm 0.22)\times 10^{-3}$ & N/A \\ \hline
        $\left<\mathbb{V}(\Delta\phi)_b / \mathbb{V}(\Delta\phi)_c \right>$ & N/A & $0.476 \pm 0.053$ & $0.642 \pm 0.060$ & N/A \\ \hline
        $\alpha_{\text{bulk}}$ & N/A & $0.37\pm 0.02$ & $0.51 \pm 0.03$ & N/A \\ \hline
        $\alpha_{\text{ctl}}$ & N/A & $0.37\pm 0.03$ & $0.48\pm 0.02$ & N/A \\ \hline
        $\tau^{ac}_{\text{bulk}}$ & N/A & $0.98 \pm 0.04$ & $1.62 \pm 0.07$ & $5.32 \pm 0.12$ \footnote{Autocorrelation function is fitted to an exponential function, whereas the rest is fitted to a homographic function.}, $2.89 \pm 0.03$ \footnote{Value coming from a fit of the power spectrum to a Lorentzian function} \\ \hline
        $\tau^{ac}_{\text{ctl}}$ & N/A & $1.00 \pm 0.09$ & $1.73 \pm 0.06$  & N/A \\ \hline
        $\epsilon_{\text{bulk}}$ & N/A & $0.434 \pm 0.037$ & $0.275 \pm 0.020$ & $0.141 \pm 0.034$ \\ \hline
        $\epsilon_{\text{ctl}}$ & N/A & $0.402 \pm 0.058$ & $0.299 \pm 0.024$ & N/A \\ \hline
        $T_{\text{eff,bulk}}$ & N/A & $(3.46\pm 0.11)\times 10^{-4}$ & $(5.56\pm 0.29) \times 10^{-4}$ & $(15.45\pm 6.36) \times 10^{-4}$ \\ \hline
        $T_{\text{eff,ctl}}$ & N/A & $(5.03\pm 0.19) \times 10^{-4}$ & $(8.75\pm 0.16) \times 10^{-4}$ & N/A \\ \hline
    \end{tabular}
    \caption{Synopsis of numerical values of the different observables. Notice that the average values of the ratios $\left<\nu_b / \nu_c \right>$ and $\left<\mathbb{V}(\Delta\phi)_b / \mathbb{V}(\Delta\phi)_c \right>$ are calculated directly as averages of the ratios and are therefore different from the ratio of the averages.}
    \label{tab:numvalues}
\end{table}

\end{document}